\documentclass[12pt]{article}
\usepackage{epsfig,graphicx}

\title{Magnetic test of chiral dynamics in QCD}
\author{  Yu.A.Simonov,\\ Institute of Theoretical and Experimental
Physics\\ 117118, Moscow, B.Cheremushkinskaya 25, Russia}
\date{}

\newcommand{\be}{\begin{equation}}
\newcommand{\ee}{\end{equation}}

\def\la{\mathrel{\mathpalette\fun <}}
\def\ga{\mathrel{\mathpalette\fun >}}
\def\fun#1#2{\lower3.6pt\vbox{\baselineskip0pt\lineskip.9pt
\ialign{$\mathsurround=0pt#1\hfil ##\hfil$\crcr#2\crcr\sim\crcr}}}

\newcommand{{\SD}}{\rm SD}

\newcommand{\vex}{\mbox{\boldmath${\rm x}$}}

\newcommand{\vesig}{\mbox{\boldmath${\rm \sigma}$}}

\newcommand{\vep}{\mbox{\boldmath${\rm p}$}}

\newcommand{\veA}{\mbox{\boldmath${\rm A}$}}

\newcommand{\veta}{\mbox{\boldmath${\rm \eta}$}}
\newcommand{\veB}{\mbox{\boldmath${\rm B}$}}

\newcommand{{\Mc}}{\mathcal{M}}

\newcommand{\lan}{\langle}
\newcommand{\ran}{\rangle}

\begin{document}

\maketitle
\begin{abstract}
Strong magnetic fields in the range $  eB\gg m^2_\pi $ effectively probe
internal quark structure of chiral mesons and test basic parameters of chiral
theory, such as $\lan\bar q q\ran, f_\pi$. We argue on general grounds that
$\lan\bar q q\ran$ should grow linearly with $eB$ when charged quark degrees of
freedom come into play. To make explicit estimates we extend previously
formulated chiral theory, including quark degrees of freedom, to the case of
strong magnetic fields and show that the quark condensate $|\lan \bar q
q\ran|_{u,d}$ grows quadratically with $eB$ for $eB\la 0.2$ GeV$^2$ and
linearly for higher field values. These results agree quantitatively with
recent lattice data and differ from $\chi PT$ predictions.

\end{abstract}

\section{Introduction}

Strong magnetic fields (MF) are expected  to be present in the early universe
\cite{1,2,3}, in neutron stars (magnetars) \cite{4,5}, and also in non-central
heavy-ion collisions \cite{6}. It is very interesting how MF can modify strong
interactions, deduced from QCD. In particular, it was recently shown \cite{7},
how MF influence asymptotic freedom behavior of the QCD strong coupling
constant and gluon exchange interaction. The interplay of confinement and MF in
neutral mesons was studied in \cite{8}, showing a dramatic change of masses
with growing MF.

Of a particular interest is  the influence of MF on  the  chiral symmetry
breaking (CSB) dynamics, and in a more general setting, on symmetry breaking
pattern in field  theory.

It was emphasized in \cite{9}, that MF has the tendency to strengthen the
chiral condensate, the phenomenon called there the magnetic catalysis, see
\cite{10} for a review.

The behavior of chiral condensate in MF was studied in the chiral perturbation
theory ($\chi PT)$ \cite{11,12,12b,13}, in Polyakov-Nambu-Jona-Lasinio (PNJL)
\cite{14}  and in other models \cite{15}. On the lattice the corresponding
analysis was done for quenched QCD in \cite{16,17}, in $n_f =2~ QCD $ in
\cite{18,19} and in $n_f =4$ SU(2) theory in \cite{20}. In all cases both in
models and on the lattice the chiral condensate grows with MF, however in
different way.

Recently a comprehensive analysis of this problem on the lattice with physical
quark masses was performed in \cite{21} and has shown a quadratic growth of
condensate up to $eB \sim 0.2$ GeV$^2$ and approximately linear behavior above
this value.

This  behavior contradicts the ($\chi PT)$ results, see e.g. \cite{11,13},
where linear growth was found with the slope almost twice as small as in
\cite{21}, and also contradicts PNJL quadratic growth at large $eB$ \cite{14}.

 As it was mentioned in \cite{12b}, the reasonable region of $\chi PT$
 application is for $eB \la m^2_\pi,$ where the quark condensate grows
 quadratically, in qualitative agreement with lattice data  however the
 linear behavior up to  1 GeV$^2$ is outside of  $\chi PT$
 reliable region.

Indeed,  the physical degrees of freedom in $\chi PT$ are associated with the
strucrureless Nambu-Goldstone mesons, and not with quarks  and antiquarks, the
role of the latter becomes more important with growing MF, and should be
decisive for $eB \ga \sigma$, where $\sigma =0.18$ GeV$^2$ is the string
tension. As it was shown in \cite{8} quark and antiquark are strongly attracted
to each other in the  plain perpendicular to MF, as it follows from the linear
growth of  the corresponding probability:  $|\psi_{q\bar q}(0) |^2 \sim eB$.

This phenomenon was called in \cite{29a} the magnetic focusing, and is the
origin of the strong enhancement of the hyperfine interaction in MF \cite{23*}.
Moreover, magnetic focusing also can produce a linear amplification with MF of
atomic, nuclear and hadronic reaction yields, as shown in \cite{24*}.

It seems reasonable, that some quantities in the chiral theory, like chiral
condensate, are proportional to $|\psi_{q\bar q} (0) |^2$, and therefore should
grow linearly with $eB$. Indeed, one can write the quark condensate (in the
Euclidean space-time) \be |\lan \bar q q\ran| = | \lan S_q (x,x)\ran | = \left|
\left\lan\left( \frac{1}{m+\hat D} \right)_{xx} \right\ran \right| = \left|
\left\lan\left( \frac{1}{m+\hat D} \right)_{xy} \left( \frac{1}{m-\hat D}
\right)_{yx}  (m-\hat D)\right\ran \right| \label{I1}\ee $= |\lan S_q(x,y)
S_{\bar q}(y,x) (m- \hat D)_{x} \ran |$,  and one  can   visualize the
probability amplitudes  of the $q\bar q$ emission at  the point $x$ and
absorption at the point $y$, combining into the factor $|\psi_{q\bar q} (0)
|^2$. This would bring us the linear behavior \be |\lan \bar q q\ran | \sim
|\psi_{q\bar q} (0)|^2 \sim eB, \label{I2}\ee which is not connected to any
chiral degrees of freedom. Therefore one  can expect, that in any model, which
takes into account the general structure (\ref{I1}) of the quark condensate,
one would end up with the linear growth  (\ref{I2}), and the main emphasis
should be on the exact quantitative form, i.e. on the coefficient in front of
$eB$ in (\ref{I2}). here the recent accurate  lattice analysis in \cite{21}
gives a good  check of analytic results, which will be used below.

It is the purpose of the present paper to study quark condensate in MF starting
from the basic QCD equations, derived earlier without MF in \cite{22,23,24,25}.
It was shown there, that GMOR relations and expressions for $\lan \bar q q\ran$
and $f_\pi$ can be  derived from the basic QCD quantities: string tension,
$\alpha_s$ and current quark masses in good agreement with experiment and
lattice data.

Recently  in  \cite{26} these results were extended to account for growing
current quark masses $m_q$, and in particular the dependence of $\lan \bar q
q\ran$ on $m_q$ was established to be in   agreement with  lattice data
\cite{27}.

In the present paper we follow the same line  of the formalism of
\cite{22,23,24,25}, but now adding  MF, and find the behavior of $\lan \bar q
q\ran$ for $u$ and $d$ quarks and their average with growing MF.

 As a result we  observe in the resulting dependence of the average $\lan
\bar q q\ran$ and $\lan \bar u u\ran, \lan \bar d d\ran$, the same features and
good quantitative agreement with the lattice data obtained in \cite{21}. The
physical reason for this dependence of $\lan \bar q q\ran $ on $eB$ is
clarified below in the paper.

The paper is organized as follows.  In the next section  a general derivation
of GMOR relation and expressions for $\lan \bar q q\ran$ and  $f_\pi$ are
given, in section 3 the MF dependence of basic terms is established,  in
section 4 results are discussed and prospectives are given.

\section{Effective chiral Lagrangian and the quark condensate}

We start with the situation without MF. In this case in \cite{24,25} the
effective Lagrangian for Nambu-Goldstone (NG) mesons with $q\bar q$ degrees of
freedom was obtained in the form

\be L_{ECL} =N_c~ tr \log [(\hat \partial + m_f)\hat 1 +  M\hat U].\label{1}\ee
where \be \hat U= \exp (i\gamma_5\hat \phi) \label{2}\ee and $M (x)$ in the
local limit is a  quark confining interaction, $M  (x) \sim \sigma|\vex| $ {
for } $  |\vex| \to \infty$ see Fig. 1, while  $\hat \phi= \phi_a t^a$,
$\phi_a(x)$ is the octet of NG mesons.

\begin{figure}\begin{center}
\includegraphics[width= 5cm,height=6cm,keepaspectratio=true]{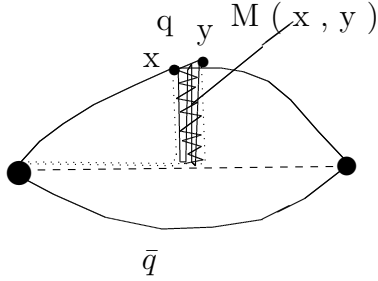}
\caption{The flux-tube operator $M(x,y)$ in the Wilson loop} \vspace{1cm}
\end{center}

\end{figure}

It is important, that $M_s (x)$, appearing at the vertices of the Green's
function in Fig.1, are finite, $$M,(0)\cong \frac{2 \sigma \lambda
}{\sqrt{\pi}}(1+O(\sigma
 \lambda^2),$$
 where $\lambda$ is the vacuum correlation length, $\lambda\cong (0.1 \div
 0.15)$ fm \cite{28}.

 Expanding (\ref{1}) to the  quadratic in $\phi$ terms, one obtains the GMOR
 relation, e.g. for the pion \be m^2_\pi f^2_\pi =\frac{\bar m_u + \bar m_d}{2}|\lan \bar \psi \psi\ran |, ~~ \bar m=\frac{m_u+m_d}{2}, \label{3}\ee
 where the quark condensate $\lan \bar \psi \psi\ran$ is\be \lan \bar \psi
 \psi\ran_M = i \lan \psi \psi^+\ran_E = - N_c tr \Lambda, ~~ \Lambda = (\hat
 \partial + \bar m + M)^{-1}.\label{4}\ee
  It was also found in \cite{23}, that  $tr \Lambda$ can be expressed via the
  $q\bar q$ Green's function $G^{(0)} (k)$
  $$ tr \Lambda = \left\lan  tr \frac{1}{M + m_q+\hat
\partial}\right\ran =\left\lan  tr\left[ \frac{1}{M
 + m_q+\hat \partial }  (M
 +
m_q-\hat \partial) \frac{1}{(M
+m_q-\hat \partial)}\right]\right\ran=$$ \be=
(M
(0) +m_q) \left\lan  tr\left[ \gamma_5 \frac{1}{M
 + m_q+ \hat \partial
}\gamma_5\frac{1}{M
 + m_q+ \hat \partial }\right] \right\ran,\label{5}\ee

and \be tr \Lambda =- (M(0) +m) G^{(0)}(k=0),\label{6}\ee where the spectral
decomposition is \cite{23}\be G^{(0)} (k) = - \sum_{n=0} \frac{c_n^2}{k^2+
m^2_n}\label{7}\ee and $c_n = \sqrt{\frac{m_n}{2}} \psi_n(0)$. Finally, the
quark condensate for $i=u,d,s$ quarks is

\be - \lan \bar q q\ran_i = N_c (M(0) + m_i) \sum^N_{n=0} \frac{\psi^2_n
 (0)}{m_n} e^{-m_n \lambda} , ~~ i = u,d,s\label{8}\ee

 Eq.   (\ref{8}) expresses
quark condensate   in terms of
 $M(0)$ and the reference spectrum in the PS channel, i.e. the spectrum, where
 chiral symmetry is not taken into account, but only confinement term $M (x)$ is
 present, the corresponding masses $m_n$, calculated in  \cite{23} are $m_0 =0.4 $
 GeV, $m_1 = 1.35$ GeV, $m_2=1.85$ GeV. Taking $\lambda =0.1 $ fm and $M(0)
 =0.15$ GeV, one obtains (cf appendix of the second paper in  \cite{23})
 \be -\frac{\lan \bar q q\ran}{n_f} = (217 {\rm MeV})^3 \frac{M(0) + \bar
 m}{(150{\rm MeV})}, ~~ f_\pi = 96 ~{\rm MeV}~\frac{M(0) + \bar
 m}{(150{\rm MeV})}.\label{9}\ee
 One can see, that our values are in the correct ballpark, close to average
 lattice and phenomenological values.

 Now we turn to the case of nonzero MF.  In this case one should replace
 $\partial_\mu \to D_\mu \equiv \partial_\mu - i e_f A_\mu^{(e)}$ and therefore
 $\Lambda_{+/-} = \frac{1}{m+M+\hat D_{+/-}}$, where
  $\hat D_{+/-}=\hat \partial_{\mp} ie_f \hat A^{(e)}$. Also one should
  introduce $\bar \Lambda_{+/-}= \frac{1}{m+M-\hat D_{+/-}}$, so that the quark
  condensate now contains contribution of both quarks and antiquarks,
  $$\lan \bar \psi\psi\ran_{M,i} = - N_c (M(0) + m_i) tr \Lambda_+ \bar
  \Lambda_-=$$
  \be - N_c (M(0) + m_i) tr (\Lambda_+ \gamma_5 \Lambda_- \gamma_5) = -
  (M(0)+m) G^{(B)} (k).\label{10}\ee

  Moreover, MF destroys both spin and isospin quantum numbers of the meson, and
  therefore one must distinguish in  the Green's function of our neutral $q\bar
  q$ system not only $(u\bar u)$ and $(d\bar d)$ components separately, but
  also the $q\bar q$ spin projections $(+-)$ and $(-+)$, since they correspond
  to different mass eigenvalues.

As a result
 Eq. (\ref{8}) in case of MF can  be rewritten for each flavor separately,
 \be |\lan \bar q q \ran_i | = N_c (M(0) + m_i) \sum^\infty_{n=0} \left(
 \frac{\frac12 |\psi^{(+-)}_{n,i} (0) |^2}{m^{(+-)}_{n,i}}+\frac{ \frac12|\psi^{(-+)}_{n,i} (0)
 |^2} {m^{(-+)}_{n,i}}\right)\label{11}\ee
 where $i=u,d,s$ and the  superscripts  $(+-)$ and $(-+)$ refer to the  quark and
 antiquark spin projections on the MF $\veB$, and the coefficients $\frac12$
 are due to $\gamma_5$ in the vertices of the Green's  function $G^{(0)}(k),
 \lan \gamma_5| \to \frac{\lan +-| - \lan -+|}{\sqrt{2}}$.

 The main problem reduces to the calculation of  the spectrum of eigenvalues
 and eigenfunctions  $\psi_n, m_n $, which are to be found from the Hamiltonian
 $H$ containing   MF $\veB$, and derived from the path integral Hamiltonian
 \cite{8}
 \be H_0= \sum^2_{i=1} \left( \frac{(\vep_i - e_i \veA^{(e)})^2}{2\omega_i}+
 \frac{m^2_i+\omega^2_i}{2\omega_i} - \frac{e_i\vesig_i
 \veB}{2\omega_i}\right)\label{12}\ee
 which can be rewritten in the form $(i=u,d,s)$
 \be H_i =\frac{1}{2\tilde \omega}\left( -\frac{d^2}{d\veta^2}
 + \left( \frac{e_i(\veta\times \veB)}{2}\right)^2 \right) + V_{\rm conf}
 (\eta) - \sum_{k=1,2} \frac{e_i  \vesig_k \veB}{2
 \omega_k},\label{13}\ee
where $\tilde \omega = \frac{\omega_1 \omega_2}{\omega_1 + \omega_2}.$

The eigenvalues $\tilde m_{n,i}$ of $H_i$ depend on $\omega_1, \omega_2$, and
the final eigenvalues $m_{n,i}$ entering in (\ref{11}) are obtained as
stationary points in variation over $ \omega_1, \omega_2$ \be\left.
\frac{\partial \tilde m_{n,i}}{\partial
\omega_{1,2}}\right|_{\omega_{1,2}^{(0)}} =0, ~~{\rm e.g.}~ \tilde
m_{n,i}^{(+-)} (\omega_1^{(0)}, \omega_2^{(0)})\equiv
m_{n,i}^{(+-)}.\label{14}\ee

To avoid purely numerical calculations, one can simplify the hamiltonian
(\ref{13}), replacing linear confinement $V_{\rm conf} (\eta) $ in (\ref{13})
by a suitable  quadratic form, with coefficient found from the stationary point
condition \be V_{conf}^{(lin)} = \sigma \eta \to V_{\rm conf}^{(\rm quadr)}
=\frac{\sigma}{2} \left( \eta^2 \gamma +\frac{1}{\gamma}\right), ~~\left.
\frac{\partial \tilde m_{n,i}}{\partial
\gamma}\right|_{\gamma=\gamma_0}=0\label{15}\ee

To check the accuracy of this replacement, one can compare the lowest mass
eigenvalues and $\psi_n(0)$ first without MF. In particular, $m_0^{\rm (quadr)}
= 2 \sqrt{3\sigma}$, while  $m_0^{\rm lin} =4\sqrt{\sigma}
\left(\frac{2.338}{3}\right)^{3/4}$, and these two figures differ by 4.5\%.

For $\psi_n(0)$ the corresponding results are $$|\psi_n (0)|^2_{\rm lin} =
\frac{0.82 \sigma^{3/2}}{4\pi} = 0.065 \sigma^{3/2};$$  \be  ~~|\psi_0
(0)|^2_{\rm quadr} = \frac{\sigma^{3/2}(\frac{c_0}{4})^{3/4}}{\pi^{3/2}}=0.065
\sigma^{3/2} (c_0 \equiv 1)\label{16}\ee

This coincidence of $|\psi_0 (0)|^2$ will be of special importance in what
follows, since the main effect of MF, as  will be seen, is the increase of
$|\psi_n (0)|^2$ in (\ref{11}) due to MF. It is clear, that  with growing $B$
the size of meson is decreasing, and the difference between $V_{\rm
conf}^{\rm(lin)}$ and $V_{\rm conf}^{(\rm quadr)}$ will be even more
suppressed, since these both interactions vanish simultaneously, $V_{\rm
conf}^{\rm (lin)} (r\to 0) =V_{\rm conf}^{(\rm quadr)} (r\to 0)=0$.

For us it will be most important how $|\psi_n(0)|^2$ depends on MF, and
especially, how MF enters in the expansion of $|\psi_n(0)|^2$ in  powers of
$B$. To this  end one can see  in the Hamiltonian (\ref{13}), that MF  enters
via the term $V_B \equiv \frac{\veta^2_\bot}{2\tilde \omega} \left( \frac{e_i
\veB}{2}\right)^2$. It is clear, that in the perturbative series expansion the
MF enters as $(e_i B)^2$ \be \psi_n(0)_B = \psi_n (0)_0 +O((e_i B)^2) + ...
\label{17}\ee

As a result of (\ref{15}) one can immediately write the analytic expressions
for $\psi_n(0)$ and $m_n$ of the  following form, e.g. \be |\psi_n(0)|^2=
\frac{1}{\pi^{3/2} r^2_\bot r_3}, ~~ r^2_\bot = 2 \left(
\left({e_qB}{~}\right)^2+ \sigma^2c\right)^{-1/2}, ~~ r_3= \left(\frac{
\sigma^2c}{4}\right)^{-1/4}\label{18}\ee where $c=\frac{4\tilde
\omega}{\gamma\sigma}$, and

\be m_{n_\bot, n_3} = \varepsilon_{n_\bot, n_3} + \frac{m^2_1+\omega^2_1- e_q
\veB\vesig_1}{2\omega_1} +\frac{m^2_2+\omega^2_2+ e_q
\veB\vesig_2}{2\omega_2}\label{19}\ee

where \be \varepsilon_{n_\bot, n_3} = \frac{1}{2\tilde{ \omega}}  [ \sqrt{e^2_q
B^2 + \sigma^2 c}(2n_\bot+1) + \sqrt{\sigma^2 c} (n_3 +\frac12)] + \frac{\gamma
\sigma}{2}\label{20}\ee

Expressions (\ref{11}),(\ref{18}), (\ref{19}), (\ref{20}) contain all
information necessary to computer quark condensate for varying MF.

\section{The MF dependence of quark  condensates}

We are interested in eigenfunctions and eigenvalues of $(+-)$ and $(-+)$ states
both for $u$ and $d$ quarks. For the $(+-)$ state, one can see in (\ref{20}),
(\ref{19}) that  at large $B$, $m_{0,n_3}^{(+-)}$ tends to a constant limit,
together with $\omega_1=\omega_2$. In this case the parameter $c$ in (\ref{18})
and (\ref{20}) which can be expressed in general as
 \be c_{+-} = \frac{4\tilde
\omega}{\gamma\sigma} = \left( \frac{a}{2}\right)^{4/3}{4/\beta}; ~~ \gamma(B)
= \beta (B) (\sigma \tilde \omega)^{-1/3}, ~~ \omega (B) = a(B)
\sqrt{\sigma}\label{44}\ee and $\beta (B), a(B)$ are changing in finite limits
and one obtains that $c_{+-}(B) \approx 1,$ for all $B$, and $m_{+-}$ tends to
a constant limit for $B\to \infty$. Hence one can write for lowest levels
$n_3=0,1,2,...$ \be |\psi^{(+-)}_{n_\bot=0, n_3} (0)|^2 \cong
\frac{\sqrt{\sigma} \sqrt{ e^2_q B^2+ \sigma^2}}{(2\pi)^{3/2}}.\label{45}\ee
Note, that in the limit $B\to 0$ this expression for $|\psi(0)|^2$ yields equal
values for $n_3=0,1,2$ as it should be for pure linear confining interaction.

For the $(-+)$ case the situation is different, and at large $B$ the stationary
point value $\omega_0^{(-+)} \approx \sqrt{2|e_q|B +\frac{\sigma}{4}}$, and the
parameter $c_{-+}$ is  increasing with $B$: \be c_{-+}(B) = \left(1+ \frac{8e_q
B}{\sigma}\right)^{2/3}.\label{46}\ee

As a result, the $|\psi_n^{(-+)}(0)|^2$ can be written as \be
|\psi_{n_3}^{(-+)}(0)|^2= (\sigma^2 c_{-+})^{3/4}
\sqrt{1+\left(\frac{e_qB}{\sigma}\right)^2\frac{1}{c_{-+}}}.\label{47}\ee

Moreover, $m^{(-+)}_{n_3} \approx 2 \sqrt{2|e_qB|+\frac{\sigma}{4}}$, and
$\frac{(\sigma^2c_{-+})^{3/4}}{m^{(-+)}_0 (B)} =\sigma$ at large $B$.

Combining (\ref{20}) and (\ref{47}) we therefore can write \be |\lan \bar q q
\ran_i (B)| = |\lan \bar q q\ran_i (0) | \frac12 \left\{
\sqrt{1+\left(\frac{e_q B}{\sigma}\right)^2}  +\sqrt{1+\left(\frac{e_q
B}{\sigma}\right)^2\frac{1}{c_{-+}}}\right\}\label{48}\ee where $c_{-+}$ is
given in (\ref{46}).

We shall be using notations of \cite{21} for the increment of quark condensate
as a function of MF (denoted $\tau_i (B)$ in \cite{9}) \be \Delta \sum_i (B) =
\frac{|\lan \bar q q\ran_i (B)|-|\lan \bar q q\ran_i (0)|}{|\lan \bar q q\ran_i
(0)|}.\label{49}\ee

The resulting values of $\Delta \sum_i (B)$ from (\ref{48}), (\ref{49}) are
compared in Table 1 with the corresponding lattice calculations in \cite{21}.

\begin{table}
\caption{Values of $\Delta \sum_i(B), i=u,d$ given by (\ref{48}) in comparison
with lattice data from \cite{21} \label{tab.1}}

\begin{center}
\begin{tabular}{|l|l|l|l|l|l|l|}
\hline\hline $eB$&0&0.2&0.4&0.6&0.8&1\\
(GeV$^2)$&&&&&&\\\hline

$\Delta \sum_u$&0&0.156&0.48&0.865&1.273&1.65\\
this paper&&&&&&\\\hline

$\Delta \sum_u$ [21]&0&0.185&0.51&0.86 &1.235&1.60\\
lattice&&&&&&\\\hline

$\Delta \sum_d$&0&0.048&0.158&0.308&0.48&0.67\\
this paper&&&&&&\\\hline

$\Delta \sum_d$ [21]&0&0.095 &0.23&0.40 &0.57&0.73\\
lattice&&&&&&\\\hline

\end{tabular}

\end{center}

\end{table}

One can see a reasonable agreement between our theory and lattice data. One can
simplify the $B$ dependence of Eq. (\ref{48}), writing \be |\lan \bar q q
\ran_i (B)| = |\lan \bar q q \ran_i (0)| \sqrt{1+
\left(\frac{e_iB}{M_i^2}\right)^2}.\label{50}\ee

 The form (\ref{50}) also satisfactorily describes data of \cite{21} and
 \cite{19}, taking $M_i$ as a fitting parameter. In the  case  of \cite{19}the fitted
 values of $M_i$ are approximately
 \be M^2_u\approx 0.35 {\rm GeV}^2, ~~ M^2_d\approx 0.27 {\rm
 GeV}^2,\label{51}\ee

 i.e. are 2 and 1.5 times larger, than $\sigma$ (the corresponding $c=c_{+-}
 =c_{-+}\cong 4$ and 4.25 for $u$ and $d$ cases), however the agreement with
 lattice data is worse.  Qualitatively the same situation (with even larger
 $M^2_i) $takes place in comparison with \cite{16, 17}.

 As a whole the behavior (\ref{48}),(\ref{50}) correctly reproduces the main
 qualitative features of the quark condensate as a function of $eB$: the quadratic behavior proportional to $(e_iB)^2$ at small $B$
 and linear behavior $\sim |e_iB|$ at large $B$.  It differs from the results of others approaches. In particular, $\chi PT$ \cite{11} predicts
  linear behavior in $eB$ with the slope much smaller, than in lattice data
 \cite{21}.

 As  it was mentioned in \cite{19}, for $\Delta \sum_u$ and $\Delta\sum_d$
 there is a simple relation
 \be \Delta \sum_u \left(\frac{B}{2}\right)=\Delta\sum_d(B),\label{52}\ee
 which is satisfied in lattice data, and, of course, is trivially satisfied in
 our definitions.

 Writing for large $B\gg \sigma$, that $\Delta \sum_i (B) \cong a_i B$, one
 immediately obtains from (\ref{52}), that $a_u =2 a_d$. This relation is
 approximately satisfied in  lattice  data \cite{19, 21}, and in our expression
 (\ref{48}), (\ref{50}).

 \section{Discussion and conclusions}

 We have used our formalism for chiral dynamics, presented in
 \cite{22,23,24,25,26}, which is derived not from purely symmetry
 considerations,  but from the QCD quark dynamics,  where
 chiral symmetry appears approximately in the small $m_q$ limit of the
 effective QCD Lagrangian. In this way all basic degrees of freedom are
 connected to the confined quarks, and it is finally the confinement, which
 dictates properties of fundamental chiral quantities $\lan \bar q q\ran$,
 $f_\pi$ etc., and  gives them numerical values, expressed via $\sigma$. This
 is in contrast with standard chiral ideas, where chiral constants are
 postulated from the beginning and chiral symmetry breaking is not connected to
 the basic QCD quark dynamics.

Actually our reasoning for the calculation of quark condensate is very simple.
After one derives scalar confining interaction $M (x)$, which acts on each
quark or antiquark, one can write quark condensate $\lan \bar q q \ran =- N_c
tr \Lambda$ as $$ tr \Lambda = \left\lan  tr \frac{1}{M  + m_q+\hat
\partial}\right\ran =\left\lan  tr\left[ \frac{1}{M  + m_q+\hat \partial }  (M +
m_q-\hat \partial) \frac{1}{(M +m_q-\hat \partial)}\right]\right\ran=$$ \be= (M
(0) +m_q) \left\lan  tr\left[ \gamma_5 \frac{1}{M  + m_q+ \hat \partial
}\gamma_5\frac{1}{M  + m_q+ \hat \partial }\right] \right\ran,\label{54}\ee
since the term with $\hat \partial$ is odd and vanishes.

The last quantity on the r.h.s. of (\ref{54}) is the $q\bar q$ Green's
function, proportional to $\psi_n^2(0)$ for each $n$ state. The only difference
in presence of MF is the replacement $\hat \partial \to (\hat \partial - ie
\hat A^{(e)})$, which immediately gives  the proportionality $\lan \bar q q\ran
\sim \psi^2 (r=0; eB)$. This latter quantity is linearly rising with $|eB|$,
since MF is ``focusing'' quark-antiquark system at small distances. Note the
similarity of (\ref{54})  and (\ref{I1}).

This phenomenon of ``magnetic focusing'' is  of  a general  character and in
the case of the chiral condensate it actually explains dynamically its growth
with $eB$, which was named before in \cite{9,10} ``magnetic catalysis''.
Recently the effect of ``magnetic focusing''  in the hyperfine interaction in
hydrogen was studied in \cite{29a}, and in the case of molecular, nuclear and
hadronic procession \cite{24*}.

 External
magnetic field is here crucial for our understanding of chiral dynamics, and
using MF one may decide, what is the role of quark dynamics in the chiral
phenomena. In this respect the comparison of the behavior of quark condensate
(or $\Delta \sum(B))$ as a function of $B$ in different models and lattice data
is showing the following:

\begin{enumerate}
    \item Standard chiral theory at large $B\sim \sigma$
    gives linear behavior (qualitatively correct) but with the wrong slope.
    \item  Our approach, Eq. (\ref{11}) predicts quadratic behavior at small
    $B$ $B\la \sigma$ and linear at larger $B$ with slopes different for $u$
    and $d$ quarks, both in agreement with existing lattice data.
    \item  The PNJL model  as shown   in \cite{21}, is also   in disagreement with
    lattice data at larger $eB$.

\end{enumerate}

Physically, it is clear, that MF, acting on quark charges, discloses the
internal quark structure of PS mesons, while the effective chiral Lagrangian,
as in  \cite{11,12} describes only internal multipionic degrees of freedom. The
latter can be important only for $B\ll 1/r^2_0$, where $r_0 \approx 0.6$ fm is
the pionic radius, i.e. for $B\ll 0.1$ GeV$^2$, while for large $B$ the
standard chiral picture is irrelevant, as it is confirmed by lattice data.

These considerations suggest the idea, that the true chiral dynamics can be
derived e.g. from the effective Lagrangian (\ref{1}), and should finally
display coexisting quark and chiral symmetric degrees of freedom, demonstrating
how the latter disappear (suppressed) for growing quark masses. How it happens
with the pionic spectra, was explained in \cite{26}, demonstrating the unifying
spectrum based on chiral and quark degrees of freedom at the same time.

The author is grateful for discussions to N.O.Agasian,  M.A.Andreichikov and
B.O.Kerbikov.

\end{document}